\documentclass[9pt,onecolumn,oneside]{osajnl}

\journal{ol} 

\setboolean{singlecolumn}{true}

\begin{document}

\title{High-power, frequency-quadrupled UV laser source resonant with the $^{1}$S$_{0}$~-~$^{3}$P$_{1}$ narrow intercombination transition of cadmium at 326.2~nm}

\author[1]{Shamaila Manzoor}
\author[1]{Jonathan N. Tinsley}
\author[1]{Satvika Bandarupally}
\author[1]{Mauro Chiarotti}
\author[1,2,3,*]{Nicola Poli}

\affil[1]{Dipartimento di Fisica e Astronomia and LENS\\Universit\`a degli Studi di Firenze, 50019 Sesto Fiorentino, Italy}
\affil[2]{Istituto Nazionale di Ottica del Consiglio Nazionale delle Ricerche\\ 50019 Sesto Fiorentino, Italy}
\affil[3]{Istituto Nazionale di Fisica Nucleare, Sezione di Firenze, 50019 Sesto Fiorentino, Italy}
\affil[*]{Corresponding author: nicola.poli@unifi.it}

\begin{abstract}
We present a novel high-power, frequency-stabilized UV laser source at 326.2~nm, resonant with the Cd $^{1}$S$_{0}$ - $^{3}$P$_{1}$ narrow intercombination transition. We have achieved a maximum produced power of 1~W at 326.2~nm by two successive frequency doubling stages of a narrow-linewidth (< 1~kHz) seed laser at 1304.8~nm. About 3.4 W of optical power at 652.4 nm is produced by a visible Raman fiber amplifier (VRFA) that amplifies and generates the second harmonic of the infrared radiation. The visible light is subsequently frequency-doubled down to 326.2~nm in a non-linear bow-tie cavity using a Brewster-cut beta-barium-borate (BBO) crystal, with a maximum conversion efficiency of around 40$\%$ for 2.5~W coupled red power. Full characterization of the laser source, together with spectroscopy signals of all Cd isotopes, spanning more than 4 GHz in the UV, are shown. 
\end{abstract}

\maketitle

\section{Introduction}
\label{sec:intro}
The lack of suitable high-power, narrow-linewidth, tunable continuous-wave laser sources in the ultraviolet (UV) - deep UV (DUV) region of the spectrum has long been a technological challenge limiting the development of experiments in atomic~\cite{Wang2016,Wilson2011,Scheid07}, molecular~\cite{Shaw2021} and optical (AMO) physics. Nonetheless, there is a great interest and motivation towards the exploration of UV and DUV transitions of atoms and molecules for a range of scientific and industrial applications. 

The narrow intercombination transitions of the transition metals Zn, Cd and Hg~\cite{Garstang1962} represent an interesting example of the viable use of UV radiation. Having an electronic structure similar to alkaline-earth atoms, due to the presence of two s-shell valence electrons, they recently attracted great attention from the atomic physics community for their potential use as optical lattice clocks~\cite{Yamanaka2015}. 
Indeed, the high-frequency UV intercombination transitions represent an attractive opportunity for high-sensitivity, large-momentum-transfer atom interferometers, due to the high $k_{eff}$ vector, and additionally provide a viable route to naturally minimize the detrimental effects of black-body radiation in optical lattice clocks~\cite{Yamanaka2015,Dzuba2019,Yamaguchi2019,Porsev2020}. Moreover, Cd with its six bosonic and two fermionic stable isotopes is an ideal candidate for precision tests of fundamental physics, as for example in the tests of the weak equivalence principle~\cite{Tinsley2022}.

Despite all these interesting features, the number of experiments tackling all the technological issues connected to the production and handling of the UV laser radiation is limited, and in the particular case of Cd, there are only a few experimental activities carried on around the world~\cite{Yamaguchi2019,Schussheim2018}. Furthermore, for the specific application of atom interferometry with Cd, both with two-photon Bragg and one-photon transitions, the requirements in terms of UV optical power are further increased, due to the general request of driving the transitions at high Rabi frequencies with homogeneous beams~\cite{Hu2019}.

Here we concentrated on addressing the $^{1}$S$_{0}$ - $^{3}$P$_{1}$ intercombination transition in Cd with a novel, high-power, narrow-linewidth, frequency-quadrupled UV laser producing up to 1~W at 326.2~nm. Early attempts to produce coherent radiation at this wavelength have been focused either on third-harmonic generation from fiber lasers up to 100~mW~\cite{Kim2008,Kim2009} or frequency doubling the output of semiconductor tapered amplifier at 650~nm, obtaining up to 50~mW in the UV~\cite{Yamaguchi2019}. More recently, the schemes for the production of high power UV at 326~nm (up to 1~W) using sum-frequency generation of two IR lasers followed by a standard frequency-doubling stage has been anticipated~\cite{Schussheim2018}. Our system takes advantage of the high power level and excellent mode quality that can be obtained from visible Raman fiber amplifiers (VRFA, MPB Communications) in the mid-infrared region (up to 10~W at 1300~nm) that integrate an efficient frequency-doubling stage provided by a periodically-poled LiNbO$_3$ waveguide (PPLN). In our case, the VRFA is fed with a frequency-stabilized external-cavity diode laser (ECDL) operating at 1304.8~nm. Subsequently, the visible light is frequency doubled in a non-linear bow-tie cavity using a Brewster-cut BBO crystal (Agile Optic)~\cite{Hannig2018}.
This approach benefits from the use of a single master laser source and a single enhancement cavity, allowing for a wider tuning range in the UV and simpler frequency stabilization down to kHz linewidths or below.

The structure of this letter is the following: in Section~\ref{sec:setup} we discuss in detail the UV laser setup; Section~\ref{characterization} discusses the spectral characterization of the laser system; Section~\ref{app} shows the implementation of the UV laser source in high-resolution spectroscopy of the intercombination transition of Cd and conclusions are reported in Section~\ref{summary}.

\section{EXPERIMENTAL SETUP}
\label{sec:setup}
\begin{figure}[ht]
\begin{center}
\includegraphics[width=.7\linewidth]{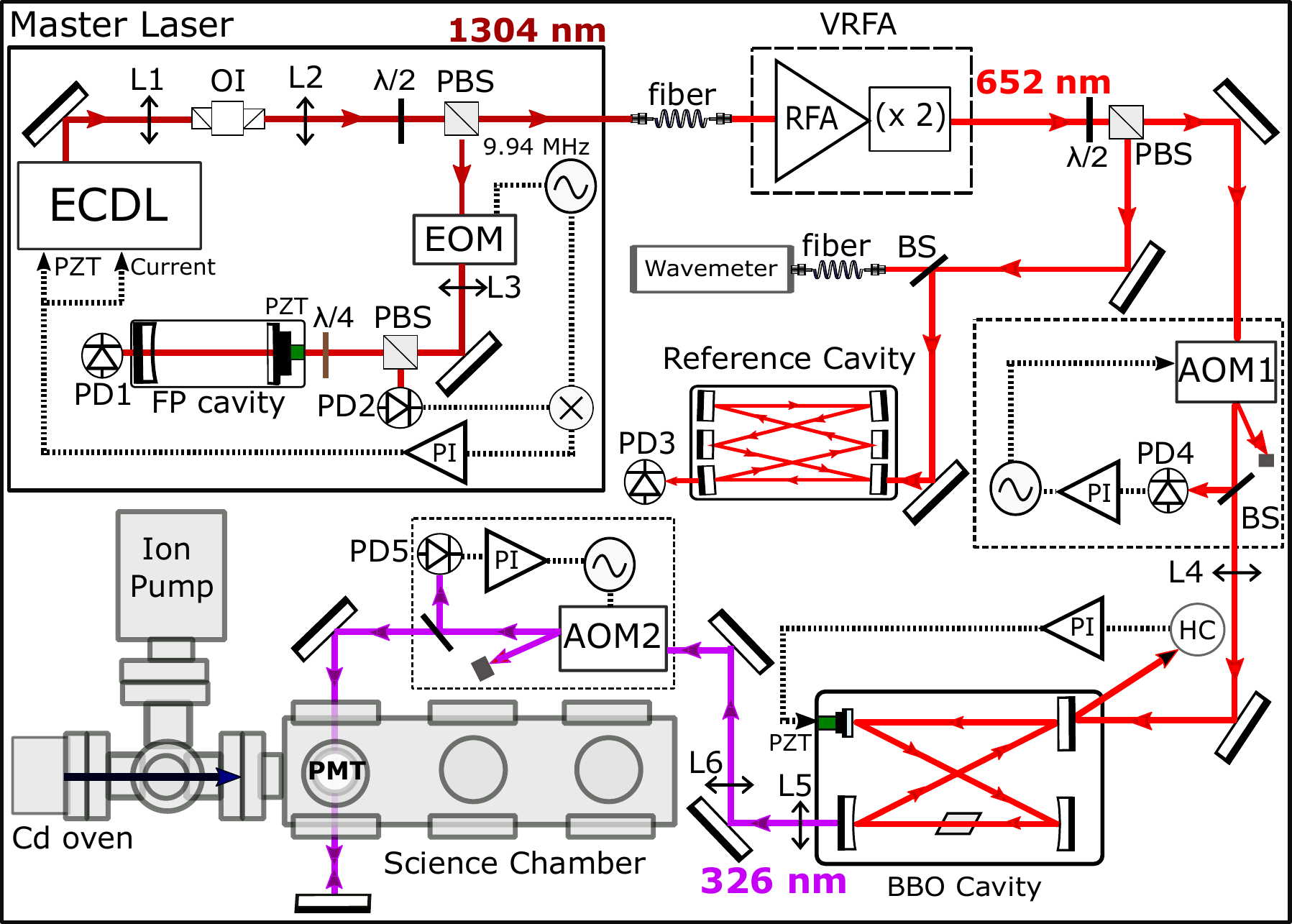}
\caption{Schematic diagram of the experimental setup used for the production 326.2~nm laser radiation and its use in spectroscopy of Cd transitions. BS beam splitter; PBS polarizing beam splitter; OI optical isolator; L(1-6) lenses; PD(1-5) photodiodes; AOM acousto-optical modulator; PI proportional-integral controller; HC H\"ansch-Couillaud locking electronics; EOM electro-optical modulator; PMT photomultiplier tube.}
\label{fig:setup}
\end{center}
\end{figure}

Fig.~\ref{fig:setup} shows the schematic diagram of the experimental setup constituted by the frequency-quadrupled UV laser and a Cd atomic beam employed for spectroscopy. The master infrared laser is a homebuilt ECDL in the Littrow configuration, based on a single-angled-facet semiconductor gain chip (Innolume) centered at 1300~nm. Due to the typical very large emission spectrum of the gain chip (60~nm), great care has been placed in selecting the proper collimation lens as a function of the external cavity geometry and gain chip parameters. Here, we choose a 0.5~NA lens (f~=~4.59~mm) to collimate the output of the laser beam onto an 800~lines/mm blazed diffraction grating. In this configuration, the laser supports single-mode operation at 1304.8~nm, delivering up to 100~mW. A 90-pm mode-hop-free continuous tuning of the emission wavelength can be obtained, without changing the cavity geometry. 

After passing through beam shaping optics and a 30~dB isolation stage, the output of the ECDL is split into two parts (see Fig.~\ref{fig:setup}). About 50~mW of light is coupled to a polarization-maintaining fiber (76~$\%$ coupling efficiency) which is then sent to the VRFA input. A secondary beam ($\sim$1~mW of power) is sent to a medium finesse ($\sim$10$^4$) Fabry-P\'{e}rot (FP) cavity for frequency stabilization of the laser using the Pound-Drever-Hall (PDH) locking method. 

The output of the Raman amplifier is directly coupled into a waveguide for the production of 652.4~nm light by a PPLN crystal in a single-pass configuration maintained at constant temperature for optimal quasi-phase-matching condition. The red beam is then sent through an acousto-optical modulator (AOM) for amplitude stabilization (see Sec.~\ref{characterization}) and subsequently mode-matched to the non-linear bow-tie cavity to produce the UV radiation~\cite{Hannig2018} with a single 500 mm lens. The non-linear cavity is formed by two half-inch diameter concave mirrors and two flat mirrors, producing a waist of 50~$\mu$m (28~$\mu$m) in the tangential (sagittal) directions between the two concave mirrors and a secondary waist of 216~$\mu$m (180~$\mu$m) between the two plane mirrors. In the position of the tight waist, a 10-mm long temperature-controlled Brewster-cut BBO crystal for type-I phase matching at 652.4~nm is placed. To optimize the phase matching and the optical input angles, the crystal is placed on a four-axis alignment stage. All the cavity mirror mounts and the crystal alignment stage are screwed directly into a monolithic aluminum housing for enhanced thermal and mechanical stability, such that all adjustment screws are accessible from the outside of the cavity housing itself~\cite{Hannig2018}. We heuristically choose a flat input coupling mirror with a reflectivity of 98.5~$\%$ to optimize the cavity impedance matching. The smaller diameter (6.7~mm) folding flat mirror is mounted on a piezoelectric transducer (PZT) to stabilize the cavity length to the seed laser by the H\"{a}nsch-Couillaud locking technique. The UV output beam from the cavity is then circularly shaped and used for high-resolution spectroscopy on a Cd atomic beam. For this application, the laser wavelength is continuously measured with a wavemeter (Bristol Instruments 621) with an accuracy of 0.2 ppm. The residual small non-linear frequency response of the laser PZT are corrected by sending part of the second-harmonic 652.4~nm light to bow-tie reference cavity with a calibrated free-spectral range (FSR) of 204.2$\pm$0.8~MHz. 

\section{LASER CHARACTERIZATION}
\label{characterization}
\begin{figure}[ht]
\begin{center}
\includegraphics[width=0.6\linewidth]{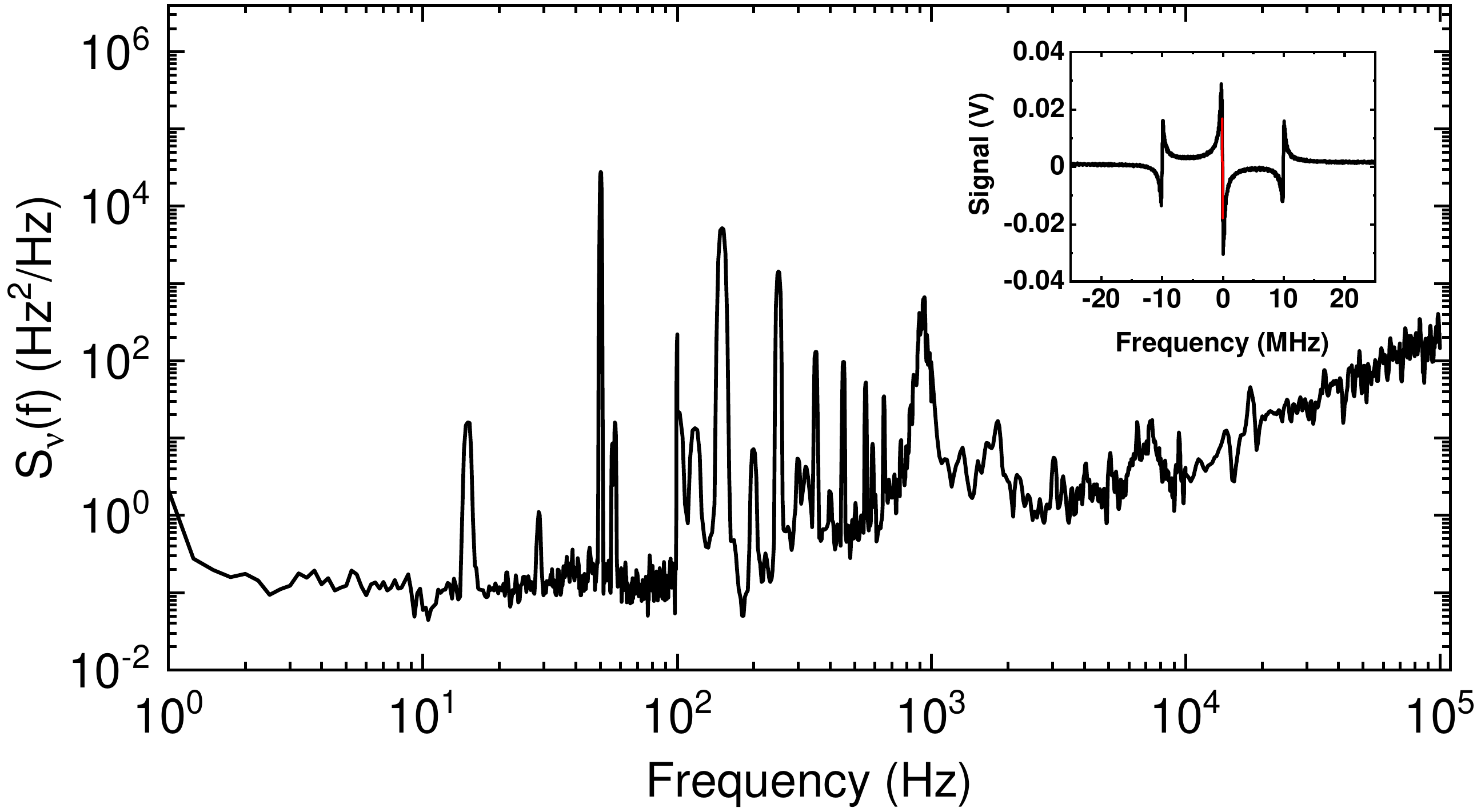}
\caption{PSD of the frequency noise of the 1304.8~nm laser in stabilized condition. The inset shows the calibrated error signal from the PDH detector obtained from the measurement of the resonance width of the cavity $\Delta\nu=290$~kHz, consistent with the $\sim$10$^4$ design finesse.}
\label{fig:Freqnoise}
\end{center}
\end{figure}

In order to effectively reduce the laser frequency noise, the ECDL has been stabilized before amplification. Indeed this scheme (Fig.\ref{fig:setup}) guarantees the highest bandwidth ($\sim$1~MHz) on the feedback to ECDL itself, which would otherwise be limited by the 330-m long fiber of the VRFA. Following this approach, the master laser linewidth has been evaluated by analyzing the power spectral density (PSD) of frequency noise $S_\nu (f)$ in locked condition, obtained by monitoring the calibrated PDH error signal on an FFT spectrum analyzer (see Fig.~\ref{fig:Freqnoise}). From the measured frequency noise PSD, by using the standard relation: $\int_{\Delta v/2}^{f_{max}} S_\phi(f) df = 1,  S_\phi(f) = S_\nu (f)/f^2$, with $f_{max}=10^5$~Hz, we estimate a fast linewidth of 0.3~kHz under locked condition. After the multiplication in the UV ($S_\phi^{UV}(f)$=16$S_\phi^{IR}(f)$), we then expect a lower limit for the linewidth of the laser emission at 326.2 nm at least an order of magnitude smaller than the natural linewidth of the Cd $^1$S$_0$~-~$^3$P$_1$ transition (2$\pi\times 67$~kHz). 

\begin{figure}[ht]
\centering
\includegraphics[width=0.7\linewidth]{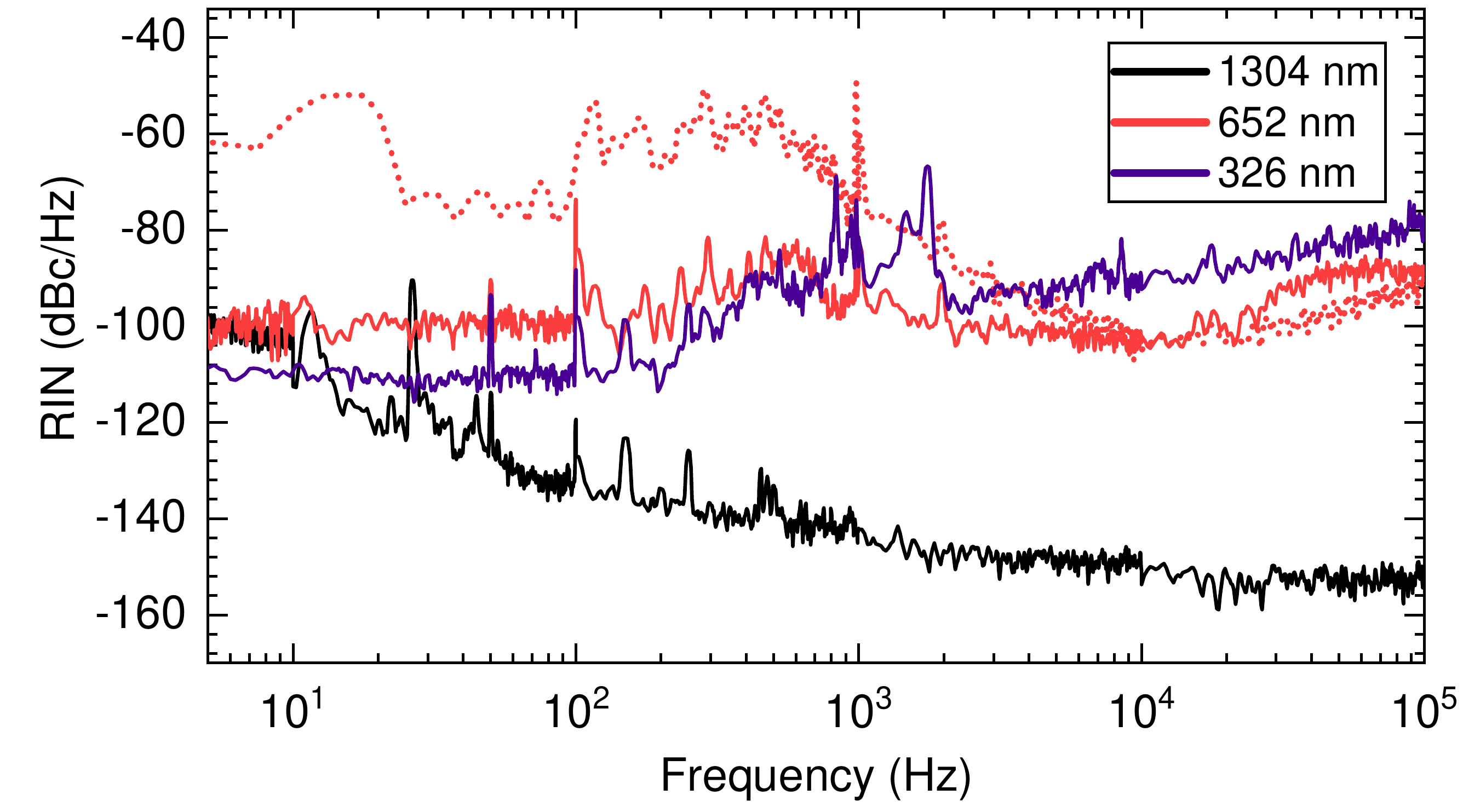}
\caption{RIN measurement of the seed laser at 1304.8~nm (black line), the SHG at 652.4~nm (red line) and UV at 326.2~nm (purple line). The SHG is shown in the stabilized (solid lines) and free-running (dotted line) cases.}
\label{fig:RIN}
\end{figure}

One drawback of using high-power VRFAs is the additional amplitude noise they typically possess and this varies from system to system. This is generally introduced by the multi-mode Yb-doped laser used to pump the silica-based Ge-doped fiber gain medium. As a consequence, the noise level of the master source is increased by several orders of magnitude in a range of frequencies not always easy to actively control~\cite{Wang2020}. 
Concerning Fig.~\ref{fig:RIN}, the relative intensity noise (RIN) is measured by monitoring the laser emission with small-area, high-bandwidth photodiodes, respectively in the IR (solid black line), VIS (solid and dotted red lines) and UV (purple line), and recording the signal on an FFT spectrum analyzer. As clearly shown in Fig.~\ref{fig:RIN}, the VFRA is adding $\sim80$~dB of noise across the 10$^2$-10$^3$~Hz region and $\sim55$~dB across the 10$^2$-10$^3$~Hz region, in the worst measured case (dotted red line vs. black line).

To reduce the detrimental effect of this noise, we implemented an active control system of the laser amplitude acting on the VIS output of the VRFA using a single AOM (AOM1 in Fig~\ref{fig:setup}). Around 90~$\%$ of the light is diffracted into the first order, which is then monitored (PD4) and intensity-stabilized via a proportional-integral controller which controls the radio-frequency level delivered to the AOM. With a maximum loop bandwidth of about $10$ kHz, a reduction of around 35~dB is achieved in the 10$^2$-10$^3$~Hz frequency band, limited by the noise floor of the electronics used. At lower frequencies ($<$10 Hz) the RIN of the VRFA output coincides with the RIN of the master IR source.

Another source of amplitude noise in our laser setup comes from  the last frequency-doubling stage in the BBO cavity down to the UV. Here, given the dependency of intracavity power from frequency detuning from cavity resonance and the intrinsic non-linear dependency of the produced UV from the fundamental power, any residual frequency noise affecting the circulating intracavity power at the fundamental wavelength, can be amplified. In our system, we observe this effect, mainly caused by additional, uncompensated acoustic and sub-acoustic noise coupled to the doubling cavity, which is not corrected by the PZT of the cavity itself. Therefore, an additional active control of the intensity of the UV beam is implemented with a second AOM (AOM2). 

We note that further optimization of the intensity stabilization system of the VRFA laser output towards higher control bandwidths up to $\sim1$~MHz is possible, as already shown elsewhere~\cite{Wang2020}. Here we instead implemented a much simpler low-frequency intensity stabilization system, eventually resulting in an integrated root mean squared intensity noise $\sigma^{uv}_{\text{rms}}$=0.6\% in the bandwidth 10-10$^4$~Hz (purple line in Fig.~\ref{fig:RIN}), sufficient for high-resolution spectroscopy of Cd transitions.

\begin{figure}[ht]
\centering
\includegraphics[width=0.7\linewidth]{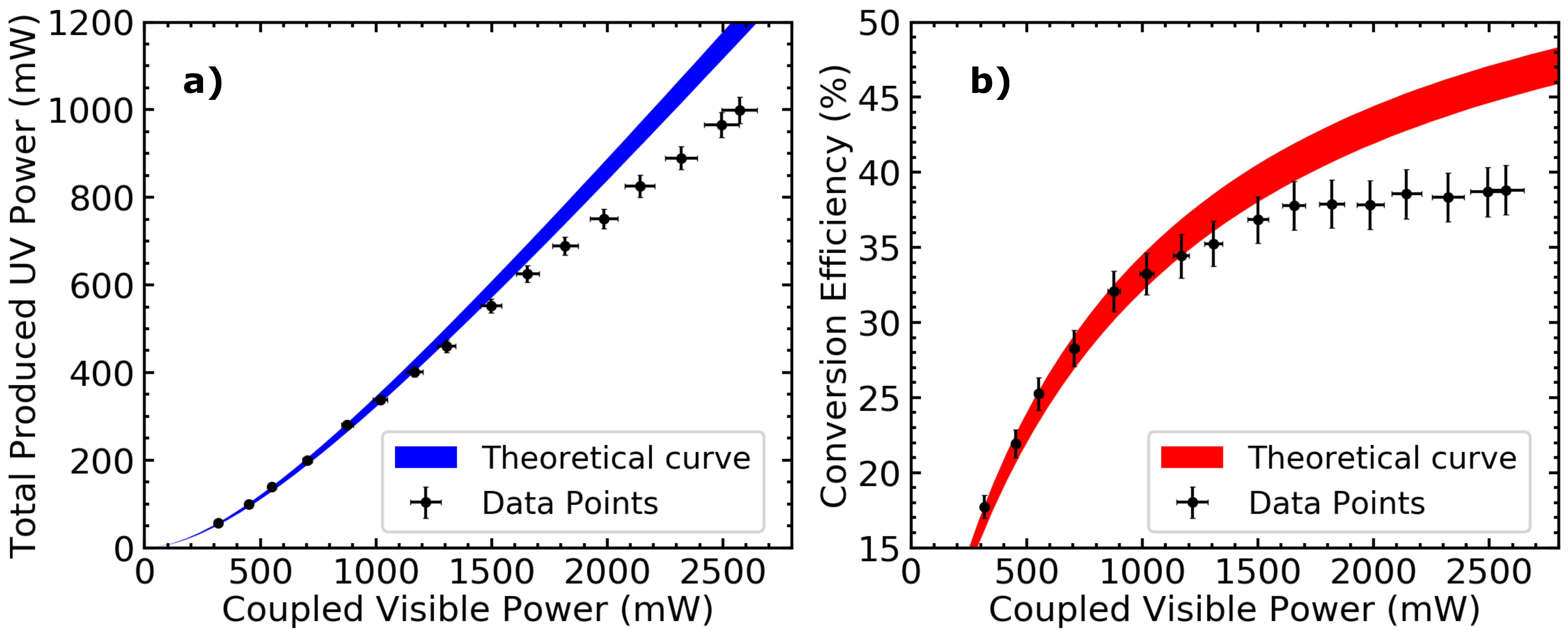}
\caption{a) Produced UV power at 326.2~nm in the BBO cavity as a function of the 652.4~nm coupled light. b) A maximum conversion efficiency of $\sim$ 40~$\%$, corresponding to 1~W of UV, is obtained. Error bars represent the systematic uncertainty of the power meter. The shaded regions in the theoretical curves represents the confidence band given by the uncertainty in the cavity finesse value.} 
\label{fig:UV}
\end{figure}

After amplitude stabilization, about 3~W of of red light is coupled (with $85\%$ efficiency) into the resonant non-linear BBO cavity with a measured finesse $F~=~225 \pm 4$. Fig.~\ref{fig:UV} shows the measured UV power and the conversion efficiency as a function of the coupled red power. For a precise determination of the total UV power produced, a calibrated UV interference filter has been used, completely removing the  leaking red light superimposed on the UV output beam. The maximum conversion efficiency of the cavity approaches 40$\%$ resulting in a giving a total UV power of 1~W for a coupled red power of about 2.5~W.
The measured power and the conversion efficiency has also been compared with a model based upon the measured finesse, cavity parameters~\cite{Boyd_1968} and BBO non-linear coefficients (blue and red regions in Fig.~\ref{fig:UV}). While at low coupled powers (below 1 W) the experimental data fits well with the model, a deviation is evident in the higher input power. We interpret this discrepancy to thermal lensing effects due to inhomogeneous temperature changes in the BBO crystal~\cite{Polzik91}.

\section{Spectroscopy of the $^1$S$_0$~-~$^3$P$_1$ intercombination transition in cadmium}
\label{app}

Here we describe an application of the developed laser to high-resolution spectroscopy of the narrow intercombination transition on a Cd atomic beam, observing signals from all eight stable Cd isotopes, demonstrating the laser's tunability over more than 4~GHz in the UV. In order to cover this large frequency scan, we apply a triangular wave signal to the PZT of the FP cavity (PZT sensitivity 22~MHz/V at 1304.8~nm). For this application in particular, we benefit from having a single, frequency-quadrupled laser, reducing the requirement on the mode-hop-free region on the master laser itself and effectively extending the laser tuning range in the UV.

The production of the Cd atomic beam is described in~\cite{Tinsley2021}, but briefly, it employs an oven filled with 10-mm long capillaries with an internal diameter of 0.23~$\mu$m to produce an atomic beam of 100~mrad divergence. At the maximum temperature of 160~$^{\circ}$C we estimate a flow rate of $\sim$10$^{12}$ atoms/s. We perform spectroscopy on the atomic beam by sending about 6~mW of linearly polarized UV light orthogonally to the atomic beam direction. The UV laser beam is expanded to a $1/e^2$ beam diameter of 5~mm to ensure a homogenous light intensity in the interaction region and a negligible transit-time broadening. A horizontal polarization has been chosen in order to maximize the fluorescence signal collected above the interaction region by a photomultiplier tube (PMT). In order to remove uncompensated residual noise in the fluorescence signal, the UV power is also monitored during the scan with a photodiode (PD5 in Fig.~\ref{fig:setup}). The atomic fluorescence collection optics consist of a lens (f~=~40~mm) and a spherical mirror with a radius of curvature of 30~mm, which are both placed inside the vacuum chamber to maximize the collected signal. The collected fluorescence is then focused on the PMT via a 2''~diameter lens (f~=~60~mm), placed outside the chamber. An optical filter with passband of 320-370 nm is attached to the PMT head to reduce the background light reaching the PMT. 

\begin{figure}[ht]
\centering\includegraphics[width=0.7\linewidth]{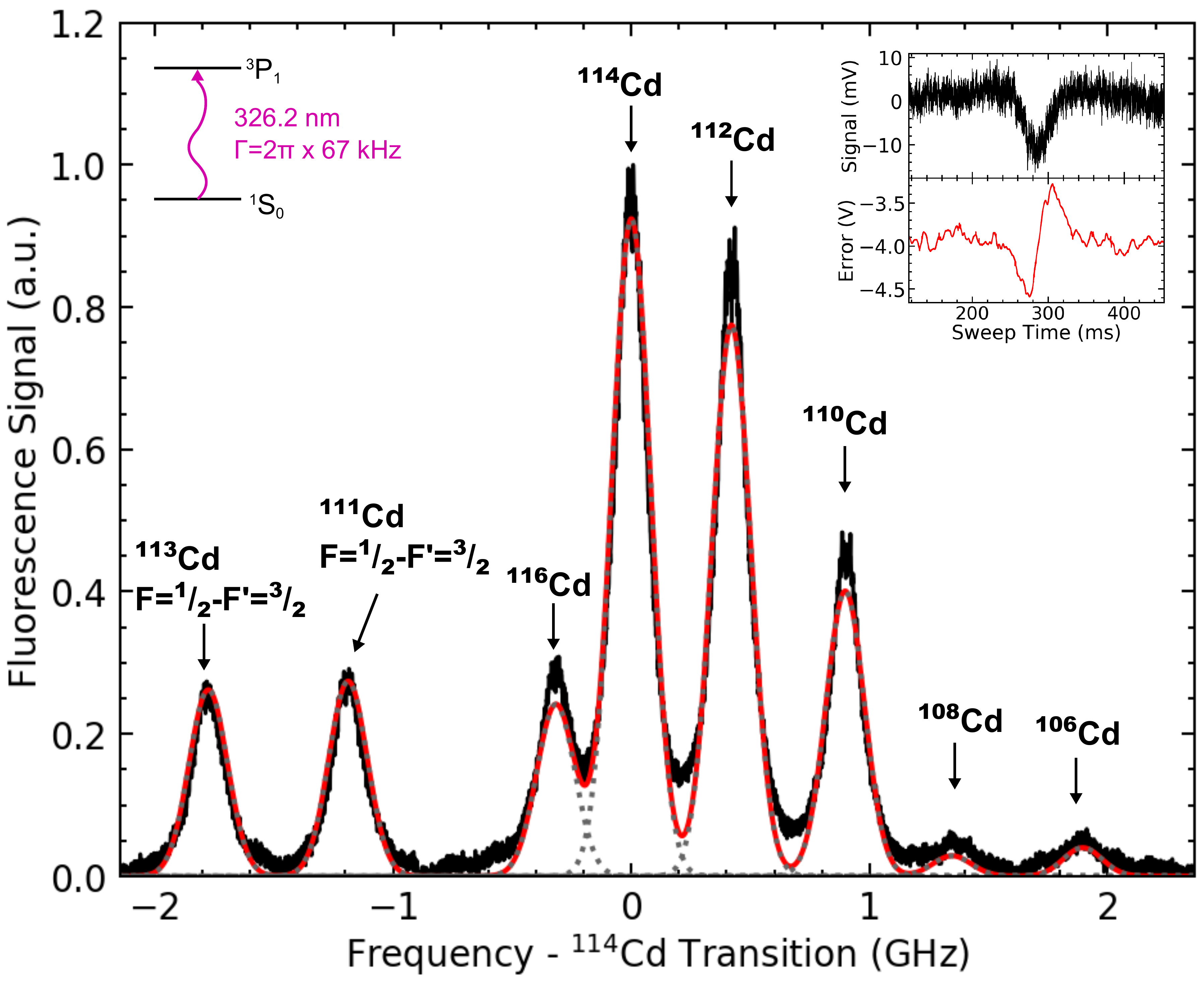}
\caption{Spectroscopic signal imaged onto the PMT. The black line shows the fluorescence of the beam, the red line the simultaneous fit to the peaks and the dotted grey lines the individual Voigt profiles. The inset shows the saturation absorption spectroscopy signal for $^{114}$Cd (top panel) and the derived error signal (bottom panel).}
\label{fig:SS}
\end{figure}

Fig.~\ref{fig:SS} shows the observed fluorescence spectrum which is fit to all the isotopes simultaneously with Voigt profiles of the same width and relative amplitudes fixed by the natural isotopic abundance and transition strengths, as previously described~\cite{Tinsley2021}.
We further performed saturation absorption spectroscopy on a Cd vapor, by sending two retro-reflected UV beams, observing the saturation dip (see inset Fig.~\ref{fig:SS}). To enable frequency stabilization onto this signal for future applications, we produced an error signal by modulating the laser frequency using AOM2 at 10~kHz and demodulating the fluorescence signal with a low-frequency lock-in amplifier.

\section{Summary and Outlook}
\label{summary}

We have presented a high-power, tunable, single-mode laser source operating at 326.2~nm, resonant with the narrow $^{1}$S$_{0}$~-~$^{3}$P$_{1}$ Cd intercombination transition. The laser produces 1~W of UV power and its operation is demonstrated by spectroscopy on a Cd atomic beam, identifying signals from all the eight isotopes of Cd, over a continuous frequency scan of more than 4~GHz. In the future, this laser source will be implemented in the laser cooling and trapping of Cd atoms using the intercombination transition~\cite{Tinsley2022}. This high-power laser source is a step towards achieving ultra-cold atoms that underpin most of the fundamental and applied research in AMO physics. Moreover, the intercombination transitions and the large number of isotopes of Cd make it an ideal candidate for searches for physics beyond the standard model through the precise determination of isotopic shifts and the study of non-linearities in King plots~\cite{Counts2020,Schelfhout2022}.

\begin{backmatter}
\bmsection{Funding}European Research Council, (772126-"TICTOCGRAV").
\bmsection{Acknowledgements} We thank Stephan Hannig and Stefan Truppe for fruitful discussions, G. Santambrogio for loaning the NIR wavemeter, Michele Sacco for technical assistance and Leonardo Salvi for initial collaboration on the development of the seed laser.
\bmsection{Disclosures} The authors declare no conflicts of interest.
\bmsection{Data availability} Data underlying the results presented in this paper are not publicly available at this time but may be obtained from the authors upon reasonable request.
\end{backmatter}

\bibliography{sample}
\bibliographyfullrefs{sample}
\end{document}